\documentclass[10pt, 
showpacs, 
preprintnumbers, 
nofootinbib, 
amsmath, 
amssymb, 
aps, 
prl, 
twocolumn, 
groupedaddress,
superscriptaddress, 
floatfix, 
showkeys
]{revtex4-2}

\usepackage{topcapt}
\usepackage[utf8]{inputenc}
\usepackage[dvipsnames]{xcolor}
\usepackage{amsfonts,amsmath,amssymb,amsthm,bm}
\usepackage{graphicx} 
\graphicspath{{paper_figures/}}
\usepackage{xspace}
\usepackage{isotope}
\usepackage{xparse}
\usepackage{aas_macros}
\usepackage[pdfpagelabels, pdfencoding=auto, psdextra]{hyperref}
\hypersetup{
 pdfsubject=Paper,
 pdfkeywords={nuclear physics} {Bayesian} {chiral EFT} {scattering},
 unicode = true,
 breaklinks = true,
 colorlinks = true,
 linkcolor = blue,
 citecolor = blue,
 menucolor = blue,
 citecolor = blue,
 urlcolor = blue
}

\newcommand{\nubb}{0 \nu \beta \beta}

\begin{document}

  \title{Statistical sensitivity of neutrinoless double-beta decay exchange mechanism discrimination by tracking experiments}
    
  \author{Jason Detwiler}
  \email{jasondet@uw.edu}
  \affiliation{Center for Experimental Nuclear Physics and Astrophysics and Department of Physics, University of Washington, Seattle, WA, USA}

  \author{Ke Han}
  \email{ke.han@sjtu.edu.cn}
  \affiliation{State Key Laboratory of Dark Matter Physics, Key Laboratory for Particle Astrophysics and Cosmology (MoE), School of Physics and Astronomy, Shanghai Jiao Tong University, Shanghai, China}
  \affiliation{Sichuan Research Institute, Shanghai Jiao Tong University, Chengdu, Sichuan, China}

  \author{Tao Li}
  \email{taoli@sjtu.edu.cn}
  \affiliation{SJTU Paris Elite Institute of Technology, Shanghai Jiao Tong University, Shanghai, China}
  \affiliation{Sichuan Research Institute, Shanghai Jiao Tong University, Chengdu, Sichuan, China}

  \date{\today}

  \begin{abstract}
Reconstruction of the individual energies and the opening angle between the electrons emitted in neutrinoless double-beta decay can probe the nature of the beyond-the-Standard-Model exchange mechanism that underlies the process.  
Although it is often stated that discrimination of the mechanism would require such measurements to be performed with high statistics, we show that this is not the case.
If a single mechanism dominates the process, its discrimination at the 1$\sigma$ level is already achieved with just a few well-reconstructed events; only $\sim$10 such events are required to reach 3$\sigma$-level discovery sensitivity. 
In the presence of realistic reconstruction uncertainties, this requirement increases to  $\sim$25 events, indicating that substantial discrimination power is retained as long as backgrounds remain small.
We conclude that the pursuit of tracking detectors for exchange-mechanism discrimination remains valuable even for ``discovery-class'' experiments in which only a few signal counts are expected.
  \end{abstract}


\maketitle

\emph{Introduction}: Intense experimental and theoretical efforts are underway to search for neutrinoless double-beta ($\nubb$) decay~\cite{Agostini:2022zub}.
Its discovery would constitute the first laboratory observation of matter created without accompanying antimatter, demonstrating that lepton number ($L$) and baryon-minus-lepton number ($B-L$) are not conserved in nature.
Such a discovery would be truly groundbreaking.

Additional strong motivation for these searches arises from their ability to probe the potential Majorana nature of the neutrino.
$L$-violating physics involving the Standard Model fields can be characterized by effective operators of odd dimensions, starting at dimension five~\cite{Cirigliano:2018yza}.
One thus naively expects $L$ violation to manifest first at dimension 5, in the unique ``Weinberg operator'' that generates a Majorana mass for the neutrino~\cite{Weinberg:1979sa}.
Under this assumption, the $\nubb$ decay rate would be dominated by the exchange of Majorana neutrinos.

But this is not the only possibility.
The Weinberg operator could vanish or be so small as to contribute negligibly to the $\nubb$ decay rate relative to operators of dimension 7 or higher.
The fact that the $\nubb$ decay diagram can be rearranged into a Majorana mass term for the neutrino~\cite{Schechter:1981bd} does not eliminate this possibility, because that contribution is exceedingly small~\cite{Duerr:2011zd}.
Thus, $\nubb$ decay could be dominated by short-range or higher-dimensional lepton-number-violating physics even when the light-neutrino Majorana-mass contribution is phenomenologically negligible.

To decide whether an observation of $\nubb$ decay is related to the neutrino nature or to other beyond-the-Standard-Model (BSM) physics, it is of paramount importance that a discovery of $\nubb$ decay be followed by an investigation of the exchange mechanism.
The exchange mechanism is imprinted both in $\nubb$ decay rates -- how the new physics couples differently to different nuclear systems -- and in the decay kinematics -- how the distribution of the energies, momenta, and angular momenta of the outgoing electrons is affected by the Lorentz structure of the exchange mechanism.
Since the claimed observation of $\nubb$ decay must be confirmed in multiple isotopes, the first tests of the exchange mechanism are likely to come from the pattern of measured half-lives~\cite{Agostini:2022bjh,Graf:2022lhj}. 
However, even in the case of a near-term discovery, which would signify that $\nubb$ decay rates can be measured with good statistical precision, half-life measurement comparisons will, for the foreseeable future, be plagued by large systematic uncertainties in nuclear matrix element calculations that can swamp the subtle variations in decay rates imprinted by the exchange mechanism. 

In contrast, the decay kinematics are strongly determined by the chirality of the exchange currents and couplings~\cite{Doi:1983wv}. 
Recognizing this, researchers have vigorously pursued detection technologies capable of reconstructing the individual electron energies ($E_1$, $E_2$) and opening angle ($\theta$) of the two emitted $\beta$ particles.
Representative examples include the tracking calorimeters NEMO~\cite{Arnold:2004xq} and SuperNEMO~\cite{SuperNEMO:2010wnd}; gaseous time-projection chambers (TPCs) such as NEXT~\cite{NEXT:2015wlq}, PandaX-III~\cite{Chen:2016qcd}, and AXEL~\cite{Hikida:2025puy}; and highly-pixelated solid-state detectors like Selena~\cite{Chavarria:2022hwx}.
These technologies already demonstrate efficient and accurate event reconstruction, though each carries distinct advantages and limitations.

For example, the NEMO-style detectors can reconstruct the $\nubb$ vertex with high precision~\cite{Arnold:2004xq}, but the thin source foils used in these experiments limit the deployable isotope mass. 
High-pressure gaseous TPCs can, in principle, accommodate substantially larger isotope masses—albeit with greater technical challenges than solid- or liquid-state approaches—but their decay vertex and track reconstructions rely on sophisticated algorithms, including Kalman filtering~\cite{Li:2021viv}, deconvolution and clustering techniques~\cite{NEXT:2023daz}, and deep convolutional neural networks~\cite{NEXT:2020jmz}.
In all cases, probing the exchange mechanism requires sufficient statistics to distinguish among the predicted decay kinematics distributions.
It is often assumed that, even if $\nubb$ decay is discovered by current-generation experiments, tracking detectors will not be able to meaningfully test the exchange mechanism until a far-future, high-statistics tracking experiment is realized~\cite{Agostini:2022zub,Graf:2022lhj}.

\begin{figure}[h]
    \centering
    \includegraphics[width=0.8\linewidth]{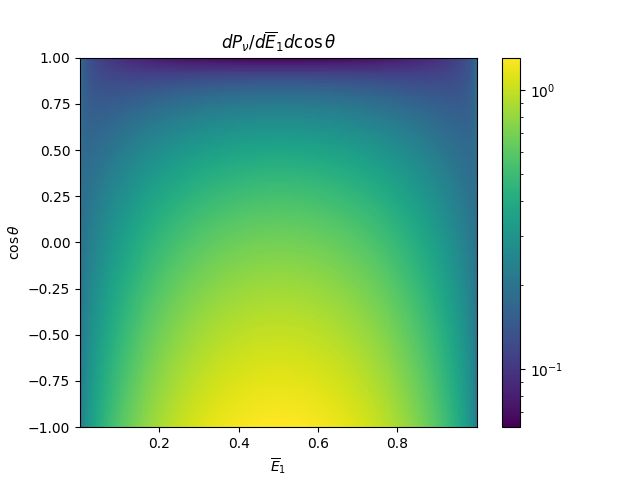}
    \includegraphics[width=0.8\linewidth]{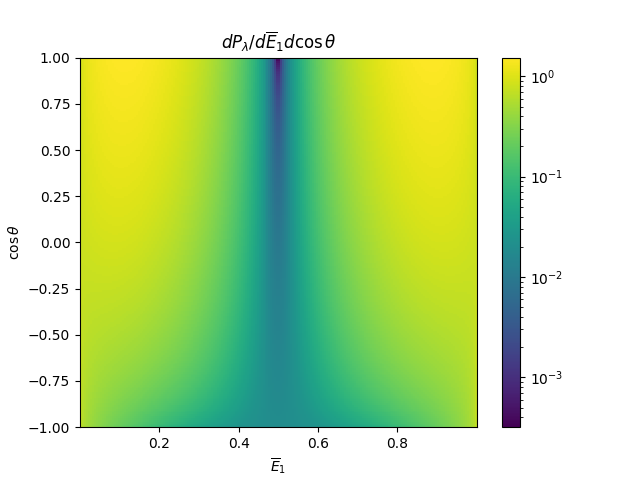}
    \includegraphics[width=0.8\linewidth]{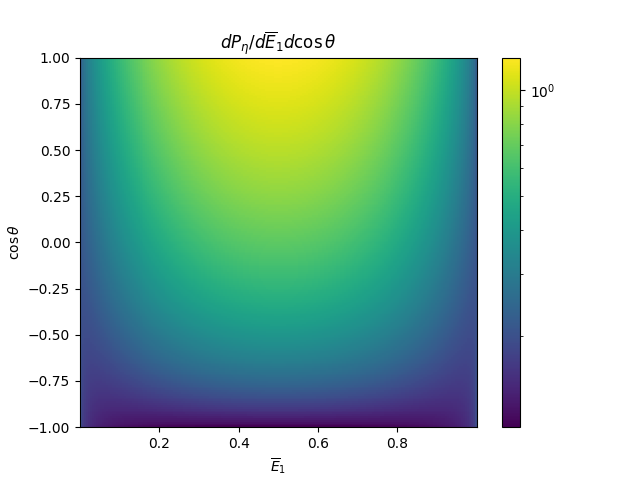}
    \caption{Ideal 2D probability distributions for $\overline{E}_1 = E_1 / Q_{\beta\beta}$ and $\cos \theta$ under the three hypotheses $H_\nu$ (top), $H_\lambda$ (middle), and $H_\eta$ (bottom).}
    \label{fig:eadists}
\end{figure}

Such pessimism is unfounded, as the space of possible decay kinematics includes predictions with pronounced differences. 
At the level of the Lagrangian, the decay kinematics are determined by how the outgoing leptons couple to the hadronic currents within the nuclear interior. 
In the Standard Model, this coupling is exclusively a left-chiral, vector-minus-axial interaction. 
By contrast, generic BSM physics can couple the left- and right-chiral hadronic currents, $J_L^\mu$ and $J_R^\mu$, to the leptonic currents in the combinations
\begin{align}
\begin{split}
\tilde{J}_L^\mu & = J_L^\mu + \chi J_R^\mu \\
\tilde{J}_R^\mu & = \eta J_L^\mu + \lambda J_R^\mu,
\end{split}
\end{align}
where $\tilde{J}_L^\mu$ and $\tilde{J}_R^\mu$ couple to the left- and right-chiral leptonic currents, respectively~\cite{Doi:1982dn}. Ordinary $\beta$ decay demonstrates that $\chi \ll 1$. We can thus identify three hypotheses representing extreme cases for $\nubb$ decay kinematics:
\begin{itemize}
\item $H_\nu$: $\lambda, \eta \ll 1$. $\nubb$ decay is dominated by light neutrino exchange, and the decay rate is determined by the effective Majorana neutrino mass $m_{\beta\beta}$. $H_\nu$ predicts $E_1$ to be peaked near $Q_{\beta\beta}/2$, with the $\beta$'s more often emitted back-to-back.
\item $H_\lambda$: $m_{\beta\beta} \rightarrow 0, ~ \lambda \gg \eta$. $\nubb$ decay is dominated by exchange mechanisms in which right-handed hadronic currents couple to right-handed leptonic currents. 
$H_\lambda$ predicts $E_1$ to be bi-modal, with the $\beta$'s more often emitted in the same direction.
\item $H_\eta$: $m_{\beta\beta} \rightarrow 0, ~ \eta \gg \lambda$. $\nubb$ decay is dominated by exchange mechanisms in which left-handed hadronic currents couple to right-handed leptonic currents. 
$H_\eta$ predicts $E_1$ to be peaked near $Q_{\beta\beta}/2$, with the $\beta$'s more often emitted in the same direction.
\end{itemize}
This qualitative picture can also be derived from a modern, effective-field-theory-oriented perspective, in which the BSM operators can be organized into three groups with distinguishable kinematics represented by the above hypotheses~\cite{Graf:2022lhj}.
For example, instances of $H_\lambda$ and $H_\eta$ are furnished by models dominated by the $C_{VR}^{(6)}$ and $C_{VL}^{(6)}$ operator groups, respectively.
Joint probability distributions for $E_1$ and $\cos \theta$ are visualized for each of these hypotheses in Fig.~\ref{fig:eadists} using $\nu$DoBe v1.0.1~\cite{Scholer:2023bnn}. In this work we always perform the $\nu$DoBe computations using nuclear shell model matrix elements for the $\nubb$ isotope $^{136}$Xe, but the results are similar for other many-body methods applied to different nuclei.

Although $\nubb$ decay exchange mechanisms typically involve linear combinations of operators from the three kinematically distinguishable operator groups, one can begin to quantify the statistical power of tracking experiments by investigating the discrimination power among the extreme hypotheses $H_\nu$, $H_\lambda$, and $H_\eta$. The differences in the distributions they predict for the kinematic observables hint that the discrimination power of tracking reconstruction is, in principle, quite high. To first order, one can divide the energy distribution into the two regimes $E_1 \approx E_2$ and $E_1 \not\approx E_2$, and the angular distributions into the acute vs.~obtuse cases, giving just four event categories. 
Each of the three hypotheses predicts that a different single category among these four is strongly represented in the data. 
Thus, it is reasonable to expect that, contrary to popular belief, significant discrimination may be provided with just a handful of events.

To explore this discrimination power quantitatively, we first consider the ideal case of a detector with a negligible background and perfect track reconstruction. 
We then explore the extent to which this discrimination power is expected to degrade for the more realistic case of a gaseous TPC detector.
We conclude with some discussion of our results and their implications.

\begin{figure}
    \centering
    \includegraphics[width=\linewidth]{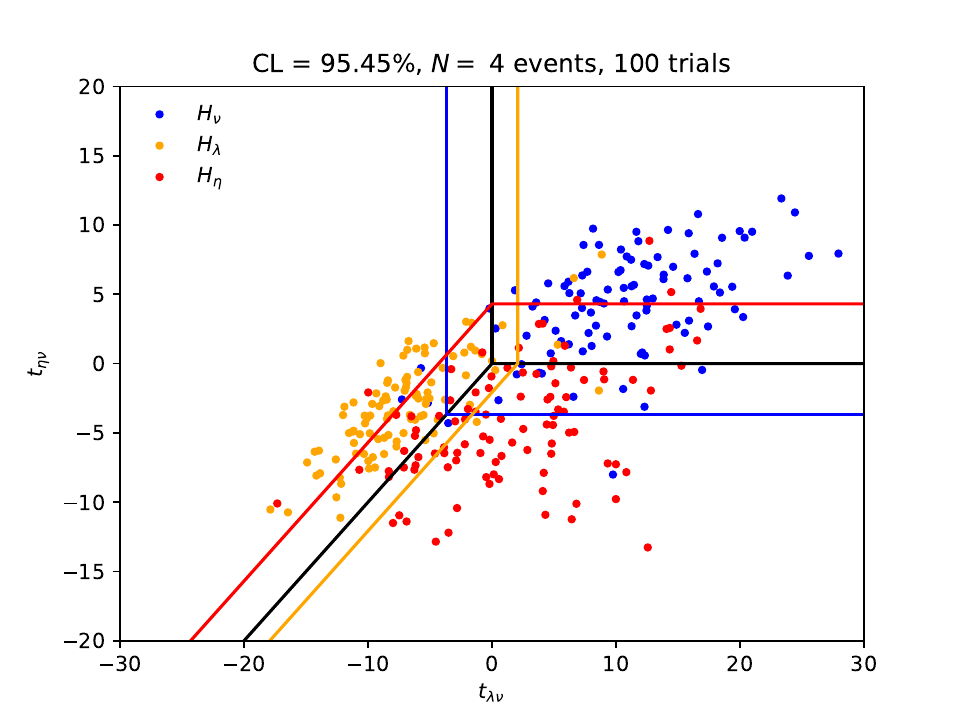}
    \caption{$t_{\lambda \nu}$ -- $t_{\eta \nu}$ plane for 100 trials of experiments observing 4 events. The black solid lines divide the plane into the three preferred regions in which one of the hypotheses is the most likely. The colored solid lines each enclose the 95.45\% inclusion region for the hypothesis corresponding to their color.}
    \label{fig:dn2LLs}
\end{figure}

\emph{Exchange mechanism tests with an ideal experiment}: As described above, we consider a theoretical landscape consisting of just the three simple hypotheses $H_\nu$, $H_\lambda$, and $H_\eta$. In the absence of background and reconstruction imperfections, the likelihood for $N$ observed events under hypothesis $H_j$ is given by
\begin{equation}
\mathcal{L}_j(\{E_{1i}, \cos \theta_i\}) = \prod_{i=1}^N \frac{dP_j(E_{1i})}{dE_1} \left(\frac{1}{2} + \frac{\alpha}{2}\cos \theta_i \right)
\end{equation}
The standard Neyman construction for multiple simple hypotheses (see 
the Appendix)
is based on logarithms of ratios of likelihoods between the different hypotheses, $t_{jk} = -2 \ln(\mathcal{L}_j/\mathcal{L}_k)$. 
In the case of three hypotheses there are only two independent $t_{jk}$, since for example one can write $t_{\lambda \eta} = t_{\lambda \nu} - t_{\eta \nu}$. The space of possible measurement outcomes can thus be visualized as single points in the $t_{\lambda \nu}$ -- $t_{\eta \nu}$ plane (see Fig.~\ref{fig:dn2LLs}). In this plane, measurement outcomes in the positive quadrant are those for which $H_\nu$ is the most likely hypothesis. The region with $t_{\lambda \nu} < 0$ and $t_{\lambda \nu} < t_{\eta \nu}$ (above the line $t_{\lambda \nu} = t_{\eta \nu}$) has $H_\lambda$ as the most likely hypothesis. $H_\eta$ is the most likely hypothesis for measurement outcomes below the $t_{\lambda \nu}$ axis and the $t_{\lambda \nu} = t_{\eta \nu}$ line.

Each hypothesis predicts a different distribution of measurement outcomes in this plane, with the distribution peaking in that hypothesis's preferred region. The Neyman construction amounts to finding acceptance regions in this plane by shifting both boundaries of a hypothesis' preferred region by the same amount $\Delta t$ to encompass a fraction of the predicted distribution that is equal to the desired confidence level (CL) of the test. When the experiment is done, one accepts each hypothesis whose acceptance region contains the measurement outcome. If the measurement outcome is near the center of the plane where all three acceptance regions overlap, all three hypotheses are accepted. If the outcome is in just one hypothesis' acceptance region, we say that that hypothesis was ``discovered'' (at the stated confidence level).

As $N$ increases, the predicted measurement outcome distributions move further and further into their preferred regions, the overlap of the acceptance regions shrinks, and the probability of identifying a single mechanism increases. 
We refer to this single-mechanism identification as “mechanism discovery,” or simply “discovery” when the context is clear.

\begin{figure}
    \centering
    \includegraphics[width=\linewidth]{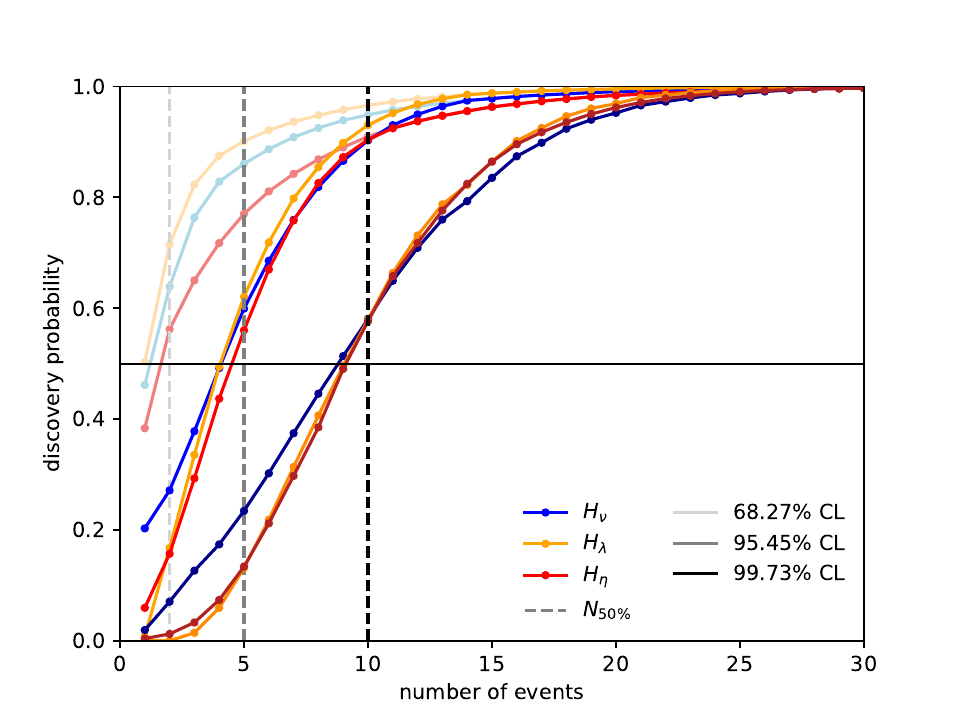}
    \caption{Discovery probability vs.~number of events for varying confidence levels. The vertical dashed lines denote $N_{50\%}$, the number of events required to reach at least 50\% discovery probability (horizontal black line).}
    \label{fig:dp}
\end{figure}

Such calculations are plotted in Fig.~\ref{fig:dp}, and we find that, as predicted by the qualitative argument at the end of the Introduction,
the growth of the mechanism discovery probability is quite rapid: it only takes a couple of events to reach strong 1$\sigma$ discovery probability, and even for a stringent 99.73\% CL, the measurement is likely to result in a discovery after the detection of just 10 events. 
It is also noteworthy that, while unlikely, the detection of just a single event can be sufficient for discovery. For example, under $H_\nu$ there's a $\sim$2\% chance for an event to occur with $E_1$ near enough to $Q_{\beta\beta}/2$ and with $\cos \theta$ near enough to $-1$ to claim a 3$\sigma$ discovery of light neutrino exchange, as such kinematics are strongly unlikely to occur under $H_\lambda$ and $H_\eta$.

\emph{Including detector effects}: Detectors have limited resolutions in energy and angular reconstruction, which affect discrimination power.  
To quantify the effect, we have constructed a conceptual high-pressure gaseous TPC within the Geant4 simulation framework~\cite{Asai:2015xno}. 
The TPC geometry resembles that of the PandaX-III experiment as described in Ref.~\cite{Li:2021viv}. 
The TPC active volume has a diameter of 1.6~m and a height of 1.2~m, containing approximately 140~kg of xenon gas at 10~bar. 
The cathode and the readout plane are positioned at the two ends of the cylindrical active volume, respectively. 

The $0\nu\beta\beta$ signals are simulated in stages.
First, the distributions of $E_1$ and $\cos \theta$ of $0\nu\beta\beta$ signals under the three hypotheses are sampled using $\nu$DoBe.
The subsequent energy-deposition processes of the two electrons in the gas medium are modeled using Geant4.
The TPC response, including electron diffusion, readout spatial resolution, and energy resolution, is simulated within the REST framework~\cite{Altenmuller:2021slh}. 
As ionization electrons drift towards the readout plane, diffusion in both transverse and longitudinal directions broadens the original tracks. 
The corresponding diffusion coefficients are set to $1.0 \times 10^{-2}$~cm$^{1/2}$ and $1.5 \times 10^{-2}$~cm$^{1/2}$, respectively, assuming xenon is mixed with a quencher gas, such as trimethylamine, to suppress diffusion. 
The readout plane is fully instrumented with pixels of $1\, \mathrm{mm} \times 1\, \mathrm{mm}$ size, which sets the spatial granularity. 
An energy resolution of 1\% Full Width at Half Maximum (FWHM) at the Q-value is assumed. 
Together, these detector effects introduce spatial and energy smearing of the event topology, making track reconstruction essential for accurately determining $E_1$ and $\cos\theta$.

We reconstruct the meandering tracks of the two $0\nu\beta\beta$ electrons using the method of Ref.~\cite{Li:2021viv}.
Events are selected within the region of interest, defined as $0.85\times$FWHM around the $Q$-value.
Hits are first grouped into a principal track and one or more subordinate tracks according to their spatial proximity.
The principal track is defined as the cluster of hits with the largest deposited energy and therefore contains the majority of the kinematic information relevant for reconstructing $E_1$ and $\cos\theta$.
After identifying the principal track, the hits belonging to it are ordered to minimize the total reconstructed path length, yielding an optimal sequence that approximates the electron's physical trajectory.
Finally, the track is represented by linearly interpolated segments between the ordered hits.
This procedure produces a continuous representation of the electron trajectory while preserving the spatial granularity determined by the TPC readout plane.
The method captures both the long, tortuous segments characteristic of multi-MeV electrons in high-pressure xenon gas and the short, localized Bragg peak near the track endpoint.
For each reconstructed event, the three-dimensional coordinates and deposited energy associated with each hit, denoted ($dx$, $dy$, $dz$, $dE$), are recorded.
These quantities form the basis for the subsequent determination of individual electron energies and the opening angle $\cos\theta$, which are the key observables for distinguishing among different $0\nu\beta\beta$ exchange mechanisms.

The electron energies and opening angles for the $0\nu\beta\beta$ events are extracted from the reconstructed tracks using a graph neural network (GNN) model based on ParticleNet~\cite{Qu:2019gqs}.
ParticleNet is an advanced architecture that operates directly on \emph{point cloud} data—an unordered, permutation-invariant set of points in an irregular spatial geometry—using dynamic graph convolutional layers to exploit local geometric correlations.
This representation is well-suited to gaseous TPC data.
In our implementation, each event is represented as a three-dimensional point cloud, where each point contains the local deposited energy at its reconstructed coordinates.
For the regression of $E_1$, ParticleNet cannot distinguish the two final-state electrons due to its inherent permutation invariance.
Therefore, instead of targeting $E_1$ directly, we train the network to predict the larger of the two individual electron energies, denoted $E_{\text{max}}$, which is uniquely defined for each event and avoids the ambiguity arising from electron exchange symmetry.

A second ParticleNet model with a similar architecture is used to reconstruct $\cos\theta$.
The input features are the same as for the $E_{\text{max}}$ regression, except that the point mask is modified to retain only the hits closest to the $0\nu\beta\beta$ decay vertex.
This focuses the network on the geometric features associated with the opening angle between the two emerging electron tracks.
Similar to~\cite{NEXT:2025olo}, we assume the decay vertex to be known, as it could be provided by auxiliary technologies such as barium tagging~\cite{NEXT:2022ita,nEXO:2018nxx} or by dedicated reconstruction techniques in future detectors.

\begin{figure*}[htbp]
    \centering
    \includegraphics[width=1.\linewidth]{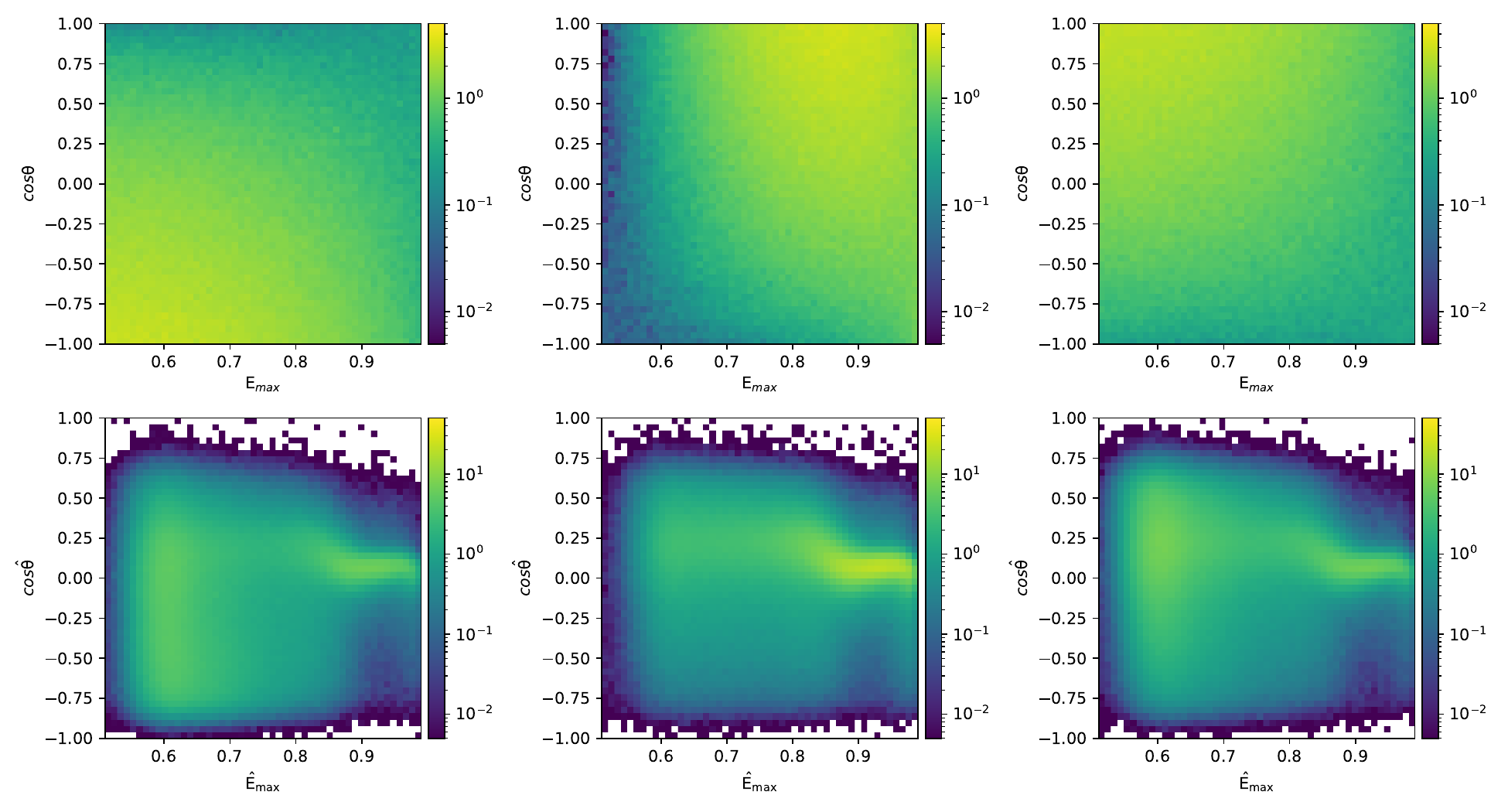}
    \caption{Reconstructed 2D probability distributions for $\hat{E}_{max} = E_{max} / Q_{\beta\beta}$ and $\cos \theta$ under the three hypotheses $H_\nu$ (left), $H_\lambda$ (middle), and $H_\eta$ (right). The true vertex is given in the $cos \theta$ reconstruction.}
    \label{fig:reconstpdf2d}
\end{figure*}

Fig.~\ref{fig:reconstpdf2d} shows the joint probability distributions of the reconstructed variables $\hat{E}_{\text{max}}$ and $\cos\hat{\theta}$ for each hypothesis.
The top row displays the input distributions of $(E_{\text{max}},\cos\theta)$, while the bottom three panels present the corresponding ParticleNet outputs for the three hypotheses.
Overall, the reconstructed distributions reproduce the main features of the inputs, particularly for the $H_\nu$ and $H_\eta$ cases.
Two limitations are evident.
First, events located near the boundaries of the input phase space are not fully reconstructed, resulting in a slightly contracted output.
Second, for $\hat{E}_{\text{max}} \gtrsim 0.8$, the $\cos\hat{\theta}$ reconstruction tends to yield values near zero, indicating a degeneracy in the ParticleNet response.
As also recognized in~\cite{NEXT:2025olo}, this behavior arises because, in this regime, the large energy asymmetry between the two electrons places the decay vertex very close to the endpoint of the higher-energy track, substantially reducing the amount of usable angular information.
Despite these effects, the reconstructed two-dimensional distributions retain clear differences among the three hypotheses, providing meaningful discriminatory power for distinguishing the underlying $0\nu\beta\beta$ exchange mechanism.

\begin{figure}[htbp]
    \centering
    \includegraphics[width=.8\linewidth]{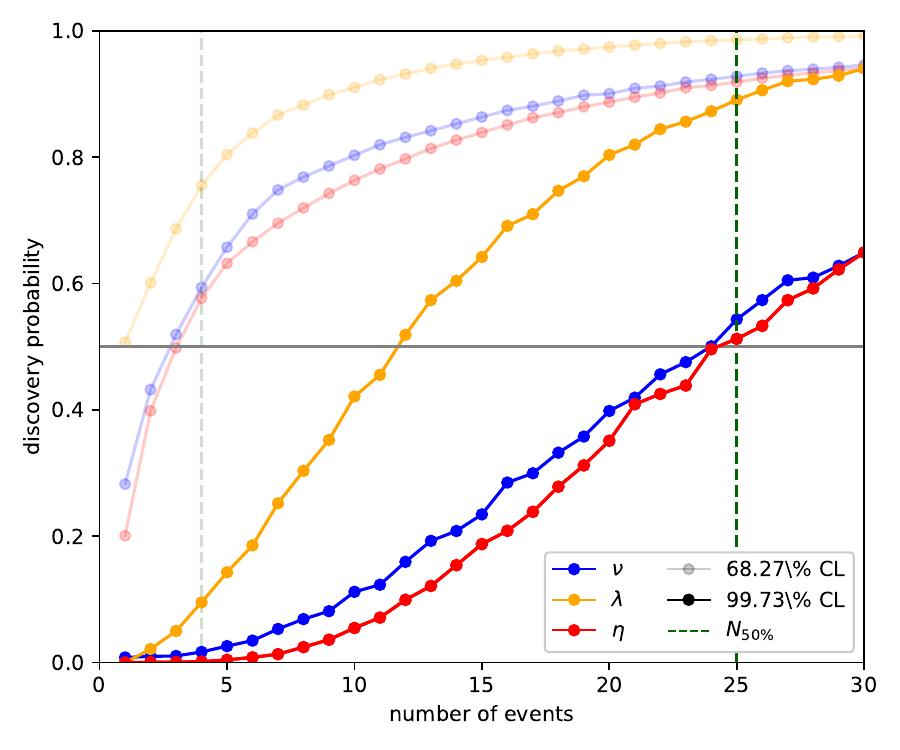}
    \caption{Discovery probability vs.~number of events for various confidence levels. The vertical dashed lines denote $N_{50\%}$, the number of events required to reach at least 50\% discovery probability (horizontal black line).}
    \label{fig:reconstdp}
\end{figure}

Using the reconstructed two-dimensional distribution of $(\hat{E}_{\text{max}}, \cos\hat{\theta})$, we repeat the discovery-probability analysis described earlier.
Fig.~\ref{fig:reconstdp} shows the resulting $3\sigma$ discovery probability curves within the ROI for our detector configuration.

For the $\lambda$ hypothesis, the discovery-probability curve exhibits only a minor shift when detector effects are included, reflecting the excellent performance of the $\hat{E}_{\text{max}}$ reconstruction.
In contrast, the curves for the $\nu$ and $\eta$ hypotheses move noticeably toward larger required event counts $N_{0\nu\beta\beta}$, indicating that detector smearing has a more substantial impact on distinguishing these scenarios that differ only in the harder-to-reconstruct angular distribution.
Taking the reconstructed distributions into account, the number of observed $0\nu\beta\beta$ events required to achieve a 50\% discovery probability at the $1\sigma$ ($3\sigma$) confidence level is found to be 4 (25).

\emph{Conclusions and discussion}: In conclusion, our results indicate that only a modest number of $0\nu\beta\beta$ events may be required to distinguish among competing underlying mechanisms.
In particular, 1$\sigma$ discrimination is already achieved after reconstructing just a few events; in the ideal case, 3$\sigma$-level discrimination requires just 10 events.

This statistical power is mostly retained even when realistic detector effects are taken into account.
Using a high-pressure gaseous TPC as a representative technology, we showed that a 100-kg-scale detector already retains substantial discriminatory power. 
The reconstructed kinematic observables, derived from ParticleNet processing of point-cloud track topologies, preserve the salient features of the underlying differential phase space.

We note that excellent reconstruction capability has recently been demonstrated by the NEXT Collaboration~\cite{NEXT:2025olo}, further underscoring the maturity of machine-learning–based topological reconstruction in gaseous TPCs.
We did not optimize for efficiency losses arising from the finite detector dimensions; these effects can be mitigated in next-generation systems. 
Future ton-scale gas TPCs—such as the proposed NEXT-1T concept~\cite{NEXT:2020amj}—will significantly reduce geometric and containment inefficiencies, thereby improving acceptance for long electron tracks and increasing the effective signal yield. The tracking performance of highly-pixelated solid-state detectors like Selena is expected to be similar to gaseous TPCs~\cite{Chavarria:2022jib}.

Tracking calorimeters, exemplified by SuperNEMO, offer intrinsically superior reconstruction of individual electron energies and directions~\cite{Krizak:2026arz}. 
The ability to localize the decay vertex to a 2D plane combined with fine-grained tracking and segmented calorimetry provides more precise measurement of the opening angle, which is one of the key observables for distinguishing right-handed-currents and exotic mechanisms from standard light-neutrino exchange. 
Based on our results, we expect that the number of required $0\nu\beta\beta$ events for mechanism discrimination in tracking calorimeters would lie between the idealized detector-free limit and the values obtained here for a gaseous TPC.

Overall, these findings highlight a realistic and achievable path toward probing the fundamental nature of lepton-number violation beyond mere discovery, demonstrating that mechanism identification may be within reach for forthcoming $0\nu\beta\beta$ experiments.

  This material is based upon work supported by the U.S.~Department of Energy, Office of Science, Office of Nuclear Physics under Award Number DE-FG02-97ER41020, the Natural Science Foundation of Shanghai, China (Grant No. 25ZR1402223), and the Natural Science Foundation of Sichuan, China (Grant No. 2026NSFSC0757).
  The authors would like to thank Lincy Aisin, Giovanni Benato, and Huilin Qu for useful discussions and suggestions.

\bibliography{refs}

\appendix
\section{Supplemental: Statistical tests for multiple simple hypotheses}
\label{app:stats}

The hypotheses $H_i$ described in this manuscript take the form of simple hypotheses, i.e., hypotheses which have no degrees of freedom and thus each specifies one distribution for the data. In the case of two simple hypotheses, a standard method is to perform the hypothesis test based on their likelihood ratio. Here we describe extensions of this method for choosing among multiple simple hypotheses. We follow the notation and terminology of~\cite{Kendall}.

Multiple simple hypotheses can be tested using a ``standard'' Neyman construction developed for compound hypotheses. The standard approach is to parameterize the compound hypothesis by a continuous parameter (or vector of parameters). But the method is applicable also to a discrete collection of simple hypotheses. Let the data $\{x_i\}$ consist of $N$ independent measurements of $x$. The likelihood for the data to be distributed according to the $j^{th}$ hypothesis $H_j$ can be written as $\mathcal{L}_j(\{x_i\}) = \prod_{i=1}^N dP_j(x_i)/dx$. The test statistic $t_j$ for $H_j$ is taken to be $-2$ times the log of a likelihood ratio:  $t_j = -2 \ln \mathcal{L}_j / \mathcal{L}_{\hat{j}}$, 
where $\hat{j}$ is the index of the hypothesis with the highest likelihood. After choosing a test level $\alpha$ (the allowable ``type-I'' error), one then determines the cumulative distributions $C_j(t)$ for each hypothesis (e.g.~via Monte Carlo methods), and computes for each a critical value $t^c_j$, taken to be the minimum value satisfying $C_j(t^c_j) \ge 1-\alpha$. One then accepts every $H_j$ (at confidence level $1-\alpha$) for which $t_j \le t^c_j$. 

A key difference between continuous and discrete compound hypotheses is the ability in the latter case for the data to isolate just one hypothesis as likely: a ``discovery'' that hypothesis $H_{\hat{j}}$ is the true hypothesis can be claimed (at confidence level $1-\alpha$) if the set of accepted hypotheses contains only $H_{\hat{j}}$. The ability to perform such a statistical discovery is related not just to the size of the test $\alpha$ but also to the distributions of the test statistic under the various hypotheses. For a given $N$, a measurement has a discovery probability for each hypothesis given by the fraction of the test statistic distribution under that hypothesis that falls outside of the acceptance regions of all other hypotheses. As $N$ increases, the overlaps of the acceptance regions shrink, and the discovery probabilities increase. Following a convention used for continuous compound hypotheses, one can state that an experiment has discovery sensitivity (at confidence level $1-\alpha$) to choose among the available simple hypotheses if $N$ is large enough to make the discovery probability greater than 50\% for all $H_j$: the measurement is ``likely'' to discover the true unknown hypothesis, whichever it is.

\end{document}